\documentstyle[epsfig]{mn2e}

\def\be{\begin{equation}}
\def\ee{\end{equation}}
\def\bea{\begin{eqnarray}}
\def\eea{\end{eqnarray}}
\def\etal{et al.~}
\def\apj{ApJ}

\title[]{GRB 060418 and 060607A: the medium surrounding the progenitor and the weak reverse shock emission}
\author[]{Z. P. Jin$^{1,2}$ and Y. Z. Fan$^{1,2,3}$\thanks{Golda Meir Fellow, E-mail: yzfan@pmo.ac.cn}\\
$^1${\sl Purple Mountain Observatory, Chinese Academy of
Science, Nanjing 210008, China}\\
$^2${\sl National Astronomical Observatories, Chinese Academy of
Sciences, Beijing 100012, China}\\
$^3${\sl The Racah Inst. of Physics, Hebrew University, Jerusalem 91904, Israel}\\
}
\date{Accepted ......
Received ......; in original form ......}

\begin{document}

\maketitle
\begin{abstract}
We constrain the circum-burst medium profile with the rise behavior
of the very early afterglow light curves of gamma-ray bursts (GRBs).
Using this method, we find a constant and low-density medium profile
for GRB 060418 and GRB 060607A, which is consistent with the
inference from the late afterglow data. In addition, we show that
the absence of the IR flashes in these two afterglows is consistent
with the standard hydrodynamical external reverse shock model.
Although a highly magnetized model can interpret the data, it is no
longer demanded. A weak reverse shock in the standard hydrodynamical
model is achievable if the typical synchrotron frequency is already
below the band at the shock crossing time.
\end{abstract}



\section{Introduction}
Quite recently, Molinari et al. (2007) reported the high-quality very early IR afterglow data
of GRB 060418 and GRB 060607A. The IR afterglow lightcurves are characterized by a sharp rise
($\sim t^3$) and then a normal decline ($\sim t^{-1.3}$), though the simultaneous X-ray
lightcurves are highly variable. The smooth joint before and after the peak time in the IR
band strongly suggests a very weak reverse shock emission.

An interesting usage of these high-quality early afterglow data is to estimate the initial
bulk Lorentz factor $\Gamma_{\rm o}$ of the outflow (Molinari et al. 2007). Such an estimate,
of course, is dependent of the circumburst medium model \cite{BM76,DL98}. For GRB 060418, a
wind profile has been ruled out by the late-time X-ray and IR afterglow data (Molinari et al.
2007). While for GRB 060607A, the X-ray data are so peculiar that the medium profile can not
be reliably determined. In this work, we use the rise behavior of the very early IR data to
pin down the density profile. This new method is valid for both bursts. We show that a
constant and low-density medium model is favored. As a result, we confirm that Molinari et
al.'s estimation of $\Gamma_{\rm o}$ for GRB 0606418 and GRB 060607A is robust.

The absence of the IR flashes for both bursts are also very
interesting. Several possible solutions for non-detection of bright
optical flashes in GRB afterglows have been discussed by Roming et
al. (2006). To account for this failing detection, it is widely
considered that the reverse shock emission would be very weak if the
outflow is highly magnetized \cite{KC84}. As shown in the numerical
calculation of Fan, Wei \& Wang (2004; Fig 1 therein), for the
magnetized reverse shock, its peak optical/IR emission increases
with $\sigma$ for $\sigma\leq 0.1$, then decreases for larger
$\sigma$, where $\sigma$ refers to the ratio of the magnetic energy
density to the particle energy density of the GRB outflow. In Fan et
al. (2004), $\sigma \leq 1$ is assumed. For highly magnetized
outflow (i.e., $\sigma> 1$), the reverse shock emission would be
suppressed further since the total electrons involved in the
emission is proportional to $1/(1+\sigma)$. Zhang \& Kobayashi
(2005) carried out such a calculation ($\sigma\leq 100$) and got
very weak reverse shock emission. In principle, the non-detection of
the optical flashes could be interpreted if $\sigma \sim 100$. This
conclusion motivated Molinari et al. (2007) to suggest these two GRB
outflows might be magnetized. However, in this work we show that for
these two bursts, the absence of the IR flashes is consistent with
the standard hydrodynamical external reverse shock model
\cite{SP99b,MR99,Kobayashi00}. Although a highly magnetized model
can interpret the data, it is no longer demanded. A weak reverse
shock in the standard hydrodynamical model is achievable if the
typical synchrotron frequency is already below the band at the shock
crossing time.

\section{Very early afterglow: constraint on the medium profile}
In this section we discuss the forward shock emission because the data show no evidence for a
dominant reverse shock component.

Firstly, we discuss a constant and low-density medium model. For $t<t_\times$, the fireball
has not been decelerated significantly by the medium, where $t_\times$ is the time when the
reverse shock crosses the outflow. The bulk Lorentz factor $\Gamma$ is thus nearly a constant,
so is the typical synchrotron frequency $\nu_m$ \cite{SPN98}. On the other hand, the maximal
specific flux $F_{\rm \nu,max} \propto N_{\rm e} \propto t^3$, where $N_{\rm e}$ is the total
number of the electrons swept by the forward shock. We thus have
\begin{equation}
F_{\rm obs} \propto F_{\rm \nu,max} (\nu_{\rm
obs}/\nu_m)^{-(p-1)/2}\propto t^3,
\end{equation}
 for $\nu_m<\nu_{\rm
obs}<\nu_c$, where $\nu_c$ is the cooling frequency and $\nu_{\rm
obs}$ is the observer's frequency. This temporal behavior is
perfectly consistent with the current data.

Secondly, we discuss a wind medium with a density profile $n_{\rm w}=3.0\times 10^{35} ~{\rm
cm^{-3}}~ A_* R^{-2}$, where $R$ is the radius of the shock front to the central engine,
$A_*=[\dot{M}/10^{-5}M_\odot~{\rm yr^{-1}}][v_w/(10^8{\rm cm}~{\rm s^{-1}})]$, $\dot{M}$ is
the mass loss rate of the progenitor, $v_w$ is the velocity of the stellar wind. Again, for
$t<t_\times$, the bulk Lorentz factor of the fireball $\Gamma$ is nearly a constant
\cite{CL00}. However, in this case, $\nu_m^{w} \propto t^{-1}$, $\nu_c^{w}\propto t$ and
$F_{\rm \nu,max}^{w}\propto t^0$, where the superscript ``$w$" represents the parameter in the
wind case. The increase of the forward shock emission can not be steeper than $t^{1/2}$, as
long as the self-absorption effect can be ignored \cite{CL00}. This temporal behavior, of
course, is inconsistent with the data. Can the synchrotron self-absorption shape the forward
shock emission significantly and then render the wind model likely? Let's examine this
possibility. In this interpretation, $\nu_a^w(t_{\rm IR,peak}) \sim 2\times 10^{14}$ Hz is
required, where $\nu_a$ is the synchrotron self-absorption frequency of the forward shock
electrons, and $t_{\rm IR,peak}$ is the peak time of the IR-band emission. In the wind model,
to get a $t^{-1.3}$ IR-band light curve, we need $p\sim 2$ and $\nu_m^w<\nu_a^w<\nu_{\rm
obs}<\nu_c^w$. Following Chevalier \& Li (2000), it is straightforward to show that $\nu_a^w
\sim 5\times 10^{13}~{\rm Hz}~\epsilon_{e,-1}^{1/3}\epsilon_{\rm
B,-2}^{1/3}A_*^{2/3}t_{d,-3}^{-1}$ and $\nu_c^w \sim 3.5\times 10^{13}~{\rm
Hz}~\epsilon_{B,-2}^{-3/2}E_{\rm
k,54}^{1/2}A_*^{-2}[(1+z)/2]^{-3/2}t_{d,-3}^{1/2}(1+Y^w)^{-2}$, where $\epsilon_e$ and
$\epsilon_B$ are the fractions of shock energy given to the electrons and magnetic field,
respectively; $z$ is the redshift, $t_d$ is the observer's time in units of day, and $E_{\rm
k}$ is the isotropic-equivalent kinetic energy of the outflow. Here and throughout this text,
the convention $Q_x=Q/10^x$ has been adopted in cgs units.

With the requirements that at $t_{\rm IR,peak}\sim 150$s, $\nu_a^w
\sim 2\times 10^{14}$ Hz and $\nu_c^w
> 2\times 10^{14}$ Hz, we have
\begin{eqnarray}
\epsilon_{e,-1}^{1/3}\epsilon_{\rm B,-2}^{1/3}A_*^{2/3} &\sim & 8,\\
\epsilon_{B,-2}^{-3/2}E_{\rm k,54}^{1/2}A_*^{-2} &> & 4(1+Y^w)^2.
\end{eqnarray}
These two relations yield $\epsilon_{B}<2\times
10^{-5}(1+Y^w)^{-4} E_{k,54}\epsilon_{e}^2$ and $A_*>160
~\epsilon_e^{-3/2}E_{k,54}^{-1/2}(1+Y^w)^2$. For such a large
contrast between $\epsilon_e$ and $\epsilon_B$, $Y^w \gg 1$. The
resulting $\epsilon_B$ and $A_*$ are too peculiar to be
acceptable.

Therefore it is very likely that the medium surrounding the GRB
progenitor has a low and constant number density. This conclusion
is also supported by the temporal and spectral analysis of the
late time X-ray and IR afterglows of GRB 060418 \cite{Molinari07}.
Here it is worth pointing out that though the multi-wavelength
afterglow modeling of many other bursts has reached a similar
conclusion \cite{PK01,Z06,Nousek06,ZhangB07,Sollerman07}, these
works were only based on the late-time afterglow data and may be
invalid for the early ones. This is because the density profile of
the circumburst medium, in principle, could vary over radius due
to the interaction between the stellar-wind and the interstellar
medium. For $R< R_c \sim {\rm several} \times 10^{16}-10^{17}$cm,
the medium may be wind-like. At larger $R$, the stalled wind
material may be ISM-like \cite{RR01}. Assuming that GRB 060418 has
such a density profile, we can estimate $R_c$ as follows. Note
that at $R_c$, the outflow has not got decelerated, which implies
that $3.8\times 10^{36}A_* \Gamma^2 m_p c^2 < E_k/2$. On the other
hand, at $t_{\rm IR,peak}$ a $\Gamma_\times \sim 200$ is likely
\cite{Molinari07}. We thus have
\begin{equation}
R_c < 2\times 10^{15}~{\rm cm}~E_{k,54}A_*^{-1}.
\end{equation}

\section{Interpreting the absence of the reverse shock emission}
\subsection{General relation between forward and reverse shock peak emission: the thin fireball case}
In the standard fireball afterglow model, there are two shocks
formed when the fireball interacts with the medium \cite{P99}, one
is the ultra-relativistic forward shock emission expanding into
the medium, the other is the reverse shock penetrating into the
GRB outflow material. The forward shock is long-lasting while the
reverse shock is short-living.  At a time $t_\times \sim \max
\{T_{90}, 60(1+z)E_{k,54}^{1/3}n_{0}^{-1/3}\Gamma_ {\rm
o,2.5}^{-8/3}\}$, the reverse shock crosses the GRB outflow, where
$n$ is the number density of the medium (please note that a dense
wind medium has been ruled out for these two bursts) and
$\Gamma_{\rm o}$ is the initial Lorentz factor of the GRB outflow.

If the fireball is thick, $t_\times \sim T_{90}$. The reverse
shock emission overlaps the prompt $\gamma-$rays and is not easy
to be detected \cite{SP99b,Kobayashi00}. In this work, we focus on
the {\it thin fireball case}, in which the peak of the reverse
shock emission and the prompt $\gamma-$rays are separated. The
relatively longer reverse shock emission renders it more easily to
be recorded by the observers. In this case,
\begin{equation}
t_\times \sim 60(1+z)~{\rm s}~E_{k,54}^{1/3}n_{0}^{-1/3}\Gamma_
{\rm o,2.5}^{-8/3}. \end{equation} After that time, the dynamics
of the forward shock can be well approximated by the
Blandford-McKee similar solution \cite{BM76}, which emission can
be estimated by
\begin{equation}
F_{\nu,{\rm max}}=6.6~{\rm mJy}~({1+z\over 2}) D_{L,28.34}^{-2}
\epsilon_{B,-2}^{1/2}E_{k,53}n_0^{1/2}, \label{eq:F_nu,max}
\end{equation}
\begin{equation}
\nu_m =2.4\times 10^{16}~{\rm Hz}~E_{\rm k,53}^{1/2}\epsilon_{\rm
B,-2}^{1/2}\epsilon_{e,-1}^2 C_p^2 ({1+z \over 2})^{1/2}
t_{d,-3}^{-3/2},\label{eq:nu_m}
\end{equation}
\begin{equation}
\nu_c = 4.4\times 10^{16}~{\rm Hz}~E_{k,
53}^{-1/2}\epsilon_{B,-2}^{-3/2}n_0^{-1}
 ({1+z \over 2})^{-1/2}t_{d,-3}^{-1/2}{1\over (1+Y)^2},
 \label{eq:nu_c}
 \end{equation}
where $p$ is the power-law index of the shocked electrons, $C_p
\equiv 13(p-2)/[3(p-1)]$, the Compton parameter $Y\sim
(-1+\sqrt{1+4\eta \epsilon_e/\epsilon_B})/2$, $\eta \sim \min\{1,
(\nu_m/\bar{\nu}_c)^{(p-2)/2} \}$ and $\bar{\nu}_c=(1+Y)^2 \nu_c$.

Following Zhang et al. (2003) and Fan \& Wei (2005), we assume that
$\epsilon_e^{\rm r}={\cal R}_{\rm e} \epsilon_{\rm e}$ and
$\epsilon_{\rm B}^{\rm r}={\cal R}_{\rm B}^2 \epsilon_{\rm B}$,
where the superscript ``$r$" represents the parameter of the reverse
shock. At $t_\times$, the reverse shock emission are governed by
\cite{FW05}
\begin{equation}
\nu_{\rm m}^{\rm r}(t_{\rm \times})={\cal R}_{\rm B}[{\cal R}_{\rm
e}(\gamma_{34,\times}-1)]^2 \nu_{\rm m} (t_{\rm
\times})/(\Gamma_{\times}-1)^2, \label{eq:nu,r_m}
\end{equation}
\begin{equation}
\nu_{\rm c}^{\rm r}(t_\times)\approx {\cal R}_{\rm B}^{-3}[(1+Y)
/(1+Y^{\rm r})]^2\nu_{\rm c}(t_\times), \label{eq:nu,r_c}
\end{equation}
\begin{equation}
F_{\rm \nu, max}^{\rm r}(t_{\rm \times})\approx \Gamma_{\rm o} {\cal
R}_{\rm B} F_{\rm \nu, max}(t_{\rm \times}), \label{eq:F,r_nu,max}
\end{equation}
where $\gamma_{34,\times}\approx (\Gamma_{\rm
o}/\Gamma_\times+\Gamma_\times/\Gamma_{\rm o})/2$ is the Lorentz
factor of the shocked ejecta relative to the initial outflow (note
that we focus on the ``thin fireball case"), $\Gamma_\times \sim
\Gamma_{\rm o}/2$ is the bulk Lorentz factor of the shocked ejecta
at $t_\times$,  $Y^{\rm r}\simeq [-1+\sqrt{1+4\eta^{\rm r}{\cal
R}_{\rm e}\epsilon_{\rm e}/({\cal R} _{\rm B}^2\epsilon_{\rm
B})}]/2$ is the Compton parameter, $\eta^{\rm r}\approx
 {\rm min}\{1,~(\nu_{\rm m}^{\rm r}/\bar{\nu}_{\rm c}^{\rm r})^{\rm (p-2)/2}\}$
 and $\bar{\nu}_c^{\rm r}=(1+Y^{\rm r})^2 \nu_c^{\rm r}$.

With the observer frequency $\nu_{\rm
 obs}$, $\nu_m^{\rm r}(t_\times)$, $\nu_c^{\rm r}(t_\times)$ and
 $F_{\rm \nu, max}^{\rm r}(t_\times)$, it is straightforward
 to estimate the peak flux of the reverse shock emission \cite{SP99}.
For the IR/optical emission (i.e., $\nu_{\rm obs}\sim 2-5\times
10^{14}$ Hz) that interests us here, we usually have $\nu_m^{\rm
r}(t_\times)<\nu_{\rm obs}<\nu_c^{\rm r}(t_\times)$. The observed
reverse shock emission is thus
\begin{eqnarray}
F^{\rm r}_{\rm obs}(t_\times) &\approx& F_{\rm \nu,max}^{\rm
r}(t_\times) [\nu_{\rm obs}/\nu_m^{\rm r}(t_\times)]^{-(p-1)/2}.
\end{eqnarray}
With eq.(\ref{eq:nu,r_m}),(\ref{eq:F,r_nu,max}) and the relation
$\Gamma_\times \sim \Gamma_{\rm o}/2$, we have \footnote{In this
work, we focus on the thin fireball case. For the thick fireball,
the reverse shock should be relativistic, we should take
$\gamma_{34}(t_\times)\approx \Gamma_{\rm o}/[2\Gamma(t_\times)]$.
Roughly, we can take $\gamma_{34}(t_\times)-1 \sim 1$. So
eq.(\ref{eq:main1}) takes a form ${F^{\rm r}_{\rm obs}(t_\times)
\over F_{\rm \nu,max}} \approx  0.4 {\cal R}_{\rm e}^{\rm p-1}
{\cal R}_{\rm B}^{\rm p+1 \over 2}[{\nu_{\rm obs} \over
\nu_m(T_{90})}]^{-(p-1)\over 2}$. So the reverse shock emission is
thus more likely to be the dominant component of the very early
afterglow (see also Zhang \& Kobayashi 2005). However, such a
reverse shock emission decays with time as $(t/T_{90})^{-2}$ or
steeper. The measurement of these very early signatures is
somewhat challenging.}
\begin{eqnarray}
{F^{\rm r}_{\rm obs}(t_\times) \over F_{\rm \nu,max}} &\approx &
\Gamma_{\rm o} {\cal R}_{\rm B}^{(p+1)\over 2} [{{\cal R}_{\rm
e}(\gamma_{34}(t_\times)-1) \over
\Gamma(t_\times)-1}]^{p-1}[{\nu_{\rm obs} \over
\nu_m(t_\times)}]^{-(p-1)\over 2}\nonumber\\
&\approx & 2^{\rm 1-p} \Gamma_{\rm o}^{\rm 2-p} {\cal R}_{\rm
e}^{\rm p-1} {\cal R}_{\rm B}^{\rm p+1 \over 2}[{\nu_{\rm obs}
\over \nu_m(t_\times)}]^{-(p-1)\over 2}\nonumber\\
&\approx & 0.08 {\cal R}_{\rm e}^{\rm p-1} {\cal R}_{\rm B}^{\rm p+1
\over 2}[{\nu_{\rm obs} \over \nu_m(t_\times)}]^{-(p-1)\over 2},
\label{eq:main1}
\end{eqnarray}
and
\begin{eqnarray}
F^{\rm r}_{\rm obs}(t_\times) \over F_{\rm obs}(t_\times) & = &
F^{\rm r}_{\rm \nu,max}[{\nu_{\rm obs} / \nu^{\rm
r}_m(t_\times)}]^{-(p-1) / 2} \over {F_{\rm \nu,max}[\nu_{\rm obs}
/ \nu_m(t_\times)]^{-(p-1)/2}}\nonumber\\
&\approx & \Gamma_{\rm o} {\cal R}_{\rm B}^{(p+1)\over 2}{\cal
R}_{\rm e}^{p-1}(\gamma_{34,\times}-1)^{p-1}
(\Gamma_{\times}-1)^{1-p}\nonumber\\
&\approx & 0.08{\cal R}_{\rm e}^{p-1}{\cal R}_{\rm B}^{p+1\over 2},
\label{eq:main2a}
\end{eqnarray}
for $\nu_{\rm m}(t_\times)\leq\nu_{\rm obs}\leq \nu_{\rm
c}(t_\times)$, or
\begin{eqnarray}
F^{\rm r}_{\rm obs}(t_\times) \over F_{\rm obs}(t_\times) & = &
F^{\rm r}_{\rm \nu,max}[{\nu_{\rm obs} / \nu^{\rm
r}_m(t_\times)}]^{-(p-1) / 2} \over {F_{\rm \nu,max}[\nu_{\rm obs}
/ \nu_m(t_\times)]^{1/3}}\nonumber\\
&\approx & 0.08{\cal R}_{\rm e}^{p-1}{\cal R}_{\rm B}^{p+1\over
2}[{\nu_{\rm obs} \over \nu_m(t_\times)}]^{-{3p-1\over 6}},
\label{eq:main2b}
\end{eqnarray}
for $\nu_{\rm c}(t_\times)>\nu_{\rm m}(t_\times)>\nu_{\rm obs}$,
here $p\sim 2.3$ and $\Gamma_{\rm o}\sim 200$ have been taken into
account. For a $p$ closer to 2, the coefficient 0.08 in
Eqs.(\ref{eq:main1}-\ref{eq:main2b}) should be lager.
Eqs.(\ref{eq:main1}-\ref{eq:main2b}) are the main result of this
paper. {\it It is now evident that to have a ${F^{\rm r}_{\rm
obs}(t_\times) \geq F_{\rm obs}(t_\times)}$, we need ${\cal
R}_{\rm e}\gg1$, or ${\cal R}_{\rm B} \gg 1$, or ${\nu_{\rm obs}
\ll \nu_m(t_\times)}$.} If ${\nu_{\rm obs} \ll
\nu_m(t_\times)}<\nu_c(t_\times)$, the forward shock will peak at
a time $t_p$ when $\nu_m(t_p)\approx
(t_p/t_\times)^{-3/2}\nu_m(t_\times) \approx \nu_{\rm obs}$. So
eq.(\ref{eq:main1}) and eq.(\ref{eq:main2b}) can be re-written as
\begin{equation}
{F^{\rm r}_{\rm obs}(t_\times) \over F_{\rm \nu,max}} = {F^{\rm
r}_{\rm obs}(t_\times) \over F_{\rm obs}(t_p)} \approx 0.08 {\cal
R}_{\rm e}^{\rm p-1} {\cal R}_{\rm B}^{\rm p+1 \over 2}({t_p \over
t_\times})^{3(p-1)\over 4}.
\end{equation}
\begin{equation}
{F^{\rm r}_{\rm obs}(t_\times) \over F_{\rm obs}(t_\times)} \approx
0.08 {\cal R}_{\rm e}^{\rm p-1} {\cal R}_{\rm B}^{\rm p+1 \over
2}({t_p \over t_\times})^{3p-1\over 4}.
\end{equation}

In the standard reverse shock model \cite{SP99b,MR99,Kobayashi00},
${\cal R}_{\rm e}={\cal R}_{\rm B}=1$. So to have a optical flash
brighter than the forward shock emission ( $F^{\rm r}_{\rm
obs}(t_\times) > F_{\rm obs}(t_p)$ ), we need
\begin{equation}
\nu_m(t_\times) > 125^{\rm 1/(p-1)} \nu_{\rm obs}~~{\rm or}~~
t_p>29^{1/(p-1)}t_\times.
\end{equation}
But to have a bright optical flash to outshine the forward shock
emission ( $F^{\rm r}_{\rm obs}(t_\times) > F_{\rm obs}(t_\times)$
), we just need
\begin{equation}
\nu_m(t_\times) > 12.5^{\rm 6/(3p-1)} \nu_{\rm obs}~~{\rm or}~~
t_p>5^{6/(3p-1)}t_\times.
\end{equation}
For typical GRB forward shock parameters $\epsilon_{e,-1}\sim
\epsilon_{B,-2}\sim E_{\rm k,53}\sim 1$ and $p\sim 2.3$, at
$t_\times \sim 100$ s, we have $\nu_m(t_\times) \sim 50 \nu_{\rm
obs}$ and $F^{\rm r}_{\rm obs}(t_\times) \sim F_{\rm \nu,max} \sim
F_{\rm obs}(t_p) \sim 4F_{\rm obs}(t_\times)$. This simple estimate
is consistent with the results of some recent/detailed numerical
calculations \cite{NP04,MKP06,Yan06}.

However, it is not clear whether these parameters, derived from modelling the late afterglow
data \cite{PK01}, are still valid for the very early afterglow data. We need high-quality
early IR/optical afterglow data to pin down this issue.

\subsection{Case study}
{\bf Analytical approach.} For GRB 060418 and GRB 060607A, their
parameters $(T_{90},~z,~E_{\gamma,52},~F_{\rm IR, peak},~t_{\rm
IR,peak})$ are $(50{\rm s},~ 1.489,~9,~50{\rm mJy},~153{\rm s})$
and $(100{\rm s},~3.082,~10,~20{\rm mJy},~180{\rm s})$,
respectively \cite{Molinari07}. Here $E_\gamma$ is the
isotropic-equivalent prompt gamma-ray energy. As shown in Molinari
\etal (2007), for $t>t_{\rm IR,peak}$, both the temporal and the
spectral data of GRB 060418 are well consistent with a
slow-cooling fireball expanding into a constant medium. To
interpret the afterglow of GRB 060607A, however, is far more
challenging. We note that the ratio between the X-ray flux and the
IR flux increases with time sharply and the late time X-ray
afterglow flux drops with time steeper than\footnote{In the jet
model, the flux declines with time as $t^{-{\rm p}}$ when we have
seen the whole ejecta. So we need a quite unusual $p\geq 4$ to
account for such a steep X-ray decline.} $t^{-4}$. These two
peculiar features, of course, can not be interpreted normally. One
speculation is that nearly all the X-ray data are the so-called
``central engine afterglow" \footnote{The central engine afterglow
emission, in principle, could be powered by either the late
internal shocks \cite{FW05,Z06,Wu07} or the late magnetic
dissipation \cite{FZP05,FP06a,GF06}. The long lasting X-ray
afterglow flat segment followed by a sharp X-ray drop has also
been detected in GRB 070110 \cite{Troja07} and is well consistent
with the emission powered by the magnetic dissipation of a
millisecond magnetar wind, as suggested by Gao \& Fan (2006) and
Fan \& Piran (2006a).} (i.e., the prompt emission of the prolonged
activity of the central engine; see Katz, Piran \& Sari [1998],
Fan, Piran \& Xu [2006]; Zhang, Liang \& Zhang [2007]) and are
independent of the IR afterglow \cite{FW05,Z06}. This kind of ad
hoc speculation is hard to be confirmed or to be ruled out.
However, the similarity between these two early IR band afterglow
light curves implies that both of them may be the forward shock
emission of a slow-cooling fireball.

Hereafter we focus on GRB 060418. The peak H-band flux is $\sim
50$ mJy, while the peak X-ray emission attributed to the forward
shock emission is likely to be $\sim 0.15$ mJy \cite{Molinari07}.
The contrast is just $\sim 300$, which suggests a $\nu_{\rm
c}(t_{\rm IR,peak})\sim 2.4\times 10^{17}$ Hz, where a
$p=2.6\pm0.1$ has been taken into account \cite{Molinari07}. On
the other hand, in the slow cooling phase, the observed flux peaks
because the observer's frequency crosses $\nu_m$ or the peak time
$ \sim t_\times$ for $\nu_m<\nu_{\rm obs}$. So we have two more
constraints: $\nu_m(t_{\rm IR,peak}) \leq 1.8\times 10^{14}$ Hz
and $F_{\rm \nu,max}\geq 50$ mJy.

With eqs.(\ref{eq:F_nu,max}-\ref{eq:nu_c}), it is straightforward
to show that
\begin{equation}
\epsilon_{B,-2}^{1/2}E_{\rm k,53} n_0^{1/2}\geq 20,
\end{equation}
\begin{equation}
\epsilon_{B,-2}^{1/2}E_{\rm k,53}^{1/2} \epsilon_{e,-1}^2 \leq
6\times 10^{-3},
\end{equation}
\begin{equation}
\epsilon_{B,-2}^{-3/2}E_{\rm k,53}^{-1/2} n_0^{-1}(1+Y)^{-2} \sim
8.
\end{equation}
These relations are satisfied with $(E_{\rm
k,53},~n_0,~\epsilon_e,~\epsilon_B,~p) \sim
(100,~1,~0.004,~0.001,~2.6)$.

Is $t_\times \sim t_{\rm IR,peak}$? The answer is positive. If
$t_\times < t_{\rm IR,peak}$,  the IR band flux will increase with
time as $t^3$ for $t \leq t_\times$ and then change with time as
$t^{1/2}$ for $t_\times <t < t_{\rm IR,peak}$ \cite{SPN98}. This
is inconsistent with the observation. So we have $t_\times \sim
t_{\rm IR,peak}>T_{90}$ and the fireball is thin. Our assumption
made in the last subsection is thus valid. Now $\nu_{\rm
m}(t_\times)\approx \nu_{\rm m}(t_{\rm IR,peak})\leq \nu_{\rm
obs}$. With eq.(\ref{eq:main1}) and ${\cal R}_{\rm e}={\cal
R}_{\rm B}=1$, we have
\begin{equation}
{F^{\rm r}_{\rm obs}(t_\times) \over F_{\rm obs}(t_\times)}\sim
{\cal O}(0.1).
\end{equation}
So the reverse shock emission is too weak to dominate over that of
the forward shock. The non-detection of the IR flashes in GRB 060418
and GRB 060607A has then been well interpreted.

{\bf Numerical fit to the afterglow of GRB 060418.} The code used
here to fit the multi-band lightcurves has been developed by Yan
\etal (2007), in which both the reverse and the forward shock
emission have been taken into account. As mentioned in the
analytical investigation, the X-ray data are flare-rich. These
flares, of course, are very hard to be understood in the external
forward shock model. Instead it may be attributed to the prolonged
activity of the central engine (Fan \& Wei 2005; Zhang \etal
2006). Assuming that the power-law decaying part (i.e., excluding
the flares) is the forward shock emission, the small contrast
between the X-ray and the H-band flux at $t_{\rm IR,peak}$
requires a $\nu_{\rm c} \sim 2.4\times 10^{17}$ Hz and thus a
small $\epsilon_{\rm B}$ and a normal $n$. The very early peak of
the IR-band afterglow light curves strongly suggests an unusual
small $\epsilon_e$. The relatively bright IR-band peak emission
implies a very large $E_{\rm k}$.

\begin{figure}
  \begin{center}
    \includegraphics{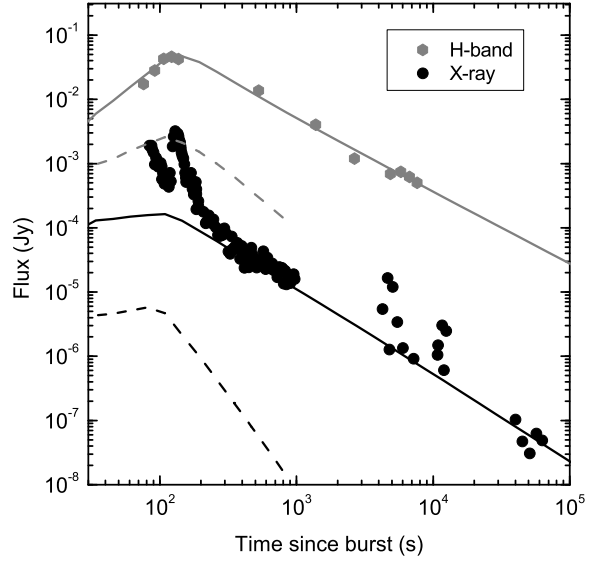}
  \end{center}
  \caption{Numerical fit to the afterglow of GRB060418. The solid and
  dashed lines represent the emission from the forward and reverse shock.
}
  \label{fig:Jin07}
\end{figure}

Our numerical results have been presented in Fig.\ref{fig:Jin07}.
The best fit parameters are  $(E_{\rm
k,53},~n_0,~\epsilon_e,~\epsilon_B,~p,~\Gamma_0) \sim
(300,~1,~0.005,~0.0002,~2.5,~600)$. The reverse shock emission is
too weak to outshine the forward shock emission (note that in this
work, ${\cal R}_{\rm e}={\cal R}_{\rm B}=1$ are assumed), as
predicted before. In the calculation, we did not take into account
the external Inverse Compton (EIC) cooling by the flare photons.
Here we discuss it analytically. Following Fan \& Piran (2006b),
the EIC cooling parameter can be estimated by $Y_{\rm EIC} \sim
0.4L_{\rm flare,48.7} \epsilon_{\rm B,-3.7}^{-1}E_{\rm
k,55.5}^{-1}{\Delta T}_3$, where $L_{\rm flare}$ is the luminosity
of the flare and $\Delta T$ is the duration of the flare. Such a
cooling correction is so small that can be ignored.

The half-opening angle $\theta_j$ of the ejecta can not be well
determined with the current data. The lack of the jet break in
H-band up to $t_d\sim 0.1$ suggests a $\theta_j>0.024$. So a
robust estimate of the intrinsic kinetic energy of GRB 060418 is
$\sim E_{\rm k}\theta_j^2/2 >8\times 10^{51}$ erg.

\section{Three kinds of reverse shock emission light curves?}
Zhang et al. (2003) suggested that there could be two types of
reverse shock emission light curves. Type-I is like that of GRB
041219a (Blake et al. 2005), in which both the forward and the
reverse shock peak emission are present [$F^{\rm r}_{\rm
obs}(t_\times)>F_{\rm obs}(t_\times)$ and $F^{\rm r}_{\rm
obs}(t_p) \sim F_{\rm obs}(t_p)$]. Type-II is like that of GRB
990123 (Akerlof et al. 1999), in which the reverse shock emission
component is so strong that outshines the peak emission of the
forward shock, i.e., $F^{\rm r}_{\rm obs}(t_p)\gg F_{\rm
obs}(t_p)$. The difference between these two kinds of reverse
shock emission has been attributed to the very different
magnetization (${\cal R}_{\rm B}$). For GRB 990123, ${\cal R}_{\rm
B} \sim 20$ (Fan et al. 2002; Zhang et al. 2003); whereas for GRB
041219a, ${\cal R}_{\rm B}\sim 3$ (Fan, Zhang \& Wei 2005).

\begin{figure}
  \begin{center}
    \includegraphics{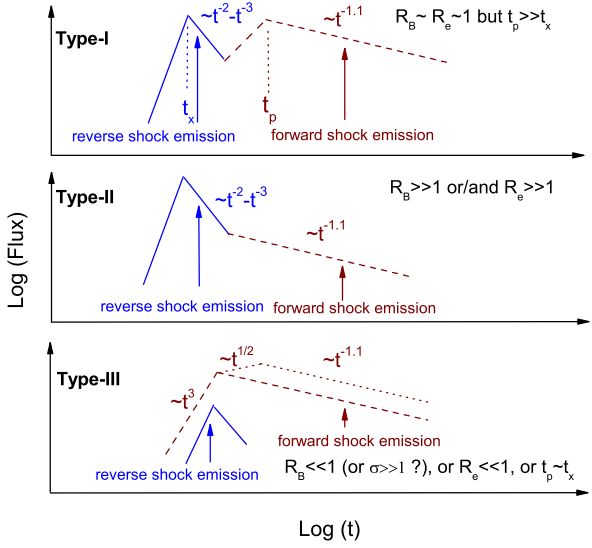}
  \end{center}
  \caption{Three types of GRB early optical afterglows. Type-I and
  Type-II were defined in Zhang et al. (2003).
  Dashed and dotted lines in Type-III represent the
  forward shock emission for $\nu_{\rm m}(t_\times)<\nu_{\rm obs}<\nu_{\rm c}(t_\times)$
  and $\nu_{\rm c}(t_\times)>\nu_{\rm m}(t_\times)>\nu_{\rm obs}$,
  respectively.
}
  \label{fig:Types}
\end{figure}

\begin{figure}
  \begin{center}
    \includegraphics[width=0.45\textwidth]{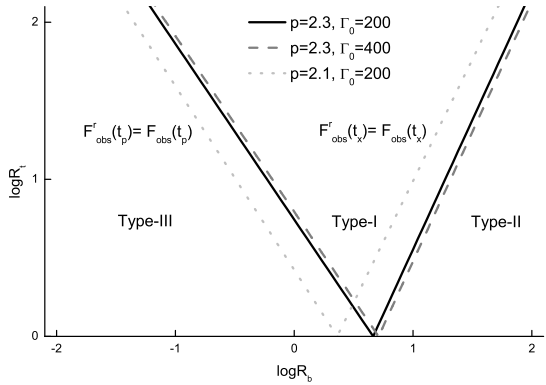}
  \end{center}
  \caption{Assuming the observer frequency located in $\nu_m^{\rm
r}(t_\times)<\nu_{\rm obs}<\nu_c^{\rm r}(t_\times)$ and $\nu_{\rm
obs}<\nu_m(t_\times)$, taking ${\cal R}_{\rm e}=1$, we identify
the regimes for the three types of GRB early optical afterglow
light curves.}
  \label{fig:Types2}
\end{figure}

GRB 060418 and GRB 060607A could be classed into Type-III, in
which the reverse shock emission is absent, i.e., $F^{\rm r}_{\rm
obs}(t_\times) \ll F_{\rm obs}(t_\times)$. We summarize these
three types of early afterglow light curves in
Fig.\ref{fig:Types}. The possible physical causes are also
presented. Fig.\ref{fig:Types2} is to identify these three types
in the ${\cal R}_{\rm B}-{\cal R}_{\rm t}$ plane, where $R_{\rm t}
\equiv t_{\rm p}/t_{\times}$.

\section{The polarimetry of prompt emission or very early afterglow: constraint
on the physical composition of GRB outflow} The polarimetry of
prompt emission or very early afterglow (i.e., the reverse shock
emission) is very important for diagnosing the composition of the
GRB outflow, since the late afterglow, taking place hours after
the trigger, is powered by the external forward shock, so that
essentially all the initial information about the ejecta is lost.

If the GRB ejecta is highly magnetized, the prompt
$\gamma-$ray/X-ray/UV/optical emission and the reverse shock
emission should be linearly polarized (Lyutikov, Pariev  \&
Blandford 2003; Granot 2003; Fan et al. 2004). This is because the
magnetic fields from the central engine are likely frozen in the
expanding shells. The toroidal magnetic field component decreases
as $r^{-1}$, while the poloidal magnetic field component decreases
as $r^{-2}$. At the typical radius for ``internal'' energy
dissipation or the reverse shock emission, the frozen-in field is
dominated by the toroidal component. For an ultra-relativistic
outflow, due to the relativistic beaming effect, only the
radiation from a very narrow cone (with the half-opening angle
$\leq 1/\Gamma$) around the line of sight can be detected. As long
as the line of sight is off the symmetric axis of the toroidal
magnetic field, the orientation of the viewed magnetic field is
nearly the same within the field of view. The synchrotron emission
from such an ordered magnetic field therefore has a preferred
polarization orientation (i.e. the direction of the toroidal
field). Consequently, the linear polarization of the synchrotron
emission of each electrons can not be averaged out and the net
emission should be highly polarized (see Fan \etal 2005a and the
references therein).

In a few events, the prompt $\gamma-$ray polarimetry are available
but the results are quite uncertain. Even for the very bright GRB
041219a, a systematic effect that could mimic the weak
polarisation signal could not be excluded \cite{McGl07}. So far,
the most reliable polarimetry is that in UV/optical band (e.g.,
Gorosabel et al. 2006). The optical polarimetry of the prompt
emission and the reverse shock emission requires a quick response
of the telescope to the GRB alert and is thus challenging. Very
recently, Mundell et al. (2007a) reported the optical polarization
of the afterglow, at 203 sec after the initial burst of
$\gamma-$rays from GRB 060418, using a ring polarimeter on the
robotic Liverpool Telescope. The percentage of polarization is
$\leq 8\%$. As shown in our section 3, the reverse shock emission
of GRB 060418 is very weak. So the optical emission at 203 sec is
mainly from the forward shock, for which a polarization $\leq 8\%$
is quite natural. The polarimetry of this particular burst thus
does not help us too much to probe the GRB outflow composition.
However, the successful polarization measurement at such an early
time demonstrates the feasibility of detecting the polarization of
prompt optical emission or reverse shock emission. {\it We expect
significant detections in the following cases}: (a) Long GRBs
followed by bright IR/optical flash, for example, GRB 990123 and
GRB 041219a \cite{Blake05}; (b) The super-long GRBs with prompt
IR/optical emission, for example, GRB 041219a \cite{Vest05}; (c)
The bright optical flare simultaneous with an X-ray flare, for
example, GRB 050904 \cite{Boer06}. The positive polarization
measurement of these events is not only a reliable diagnosis of
the physical composition of the GRB/X-ray flare outflow but also a
good probe of the Quantum Gravity (see Fan, Wei \& Xu 2007 and the
references therein).

\section{Summary}
The temporal behavior of the very early afterglow data is valuable to constrain the medium
profile surrounding the progenitor. For GRB 060418 and GRB 060607A, the sharp increase of the
very early H-band afterglow light curve has ruled out a WIND-like medium. This conclusion is
further supported by the late time X-ray and IR afterglow data of GRB 060418. This rather
robust argument is inconsistent with the canonical collapsar model, in which a dense stellar
wind medium is expected. More fruitful very early IR/optical data are needed to draw a more
general conclusion.

The absence of the reverse shock signatures in the high-quality IR
afterglows of GRB 060418 and GRB 060607A may indicate the outflows
being strongly magnetized. We, instead, show that the non-detection
of IR flashes in these two events is consistent with the standard
reverse shock model. Although a highly magnetized model can
interpret the data, it is no longer demanded. The physical reason is
that in these two bursts, $\nu_m(t_\times)\sim 2-5\times 10^{14}$
Hz. Such a small $\nu_m(t_\times)$ will influence our observation in
two respects. One is that $\nu^{\rm r}_m(t_\times)\sim
\nu_m(t_\times)/4\Gamma_{\rm o}^2 \sim 10^{9}{\rm Hz}$. The
corresponding emission in IR band is thus very weak. The other is
that now the forward shock peaks at IR/optical band at $t_\times$.
Consequently, the IR/optical emission of the reverse shock can not
dominate over that of the forward shock.

It is not clear whether the absence of the optical flashes in most
GRB afterglows \cite{Roming06} can be interpreted in this way or
not. Of course, one can always assume ${\cal R}_{\rm e}\ll 1$
or/and ${\cal R}_{\rm B}\ll1$ to solve this puzzle. But before
adopting these phenomenological approaches, one should explore the
physical processes that could give rise to these modifications.
The Poynting flux dominated outflow ($\sigma \sim 100$ after the
prompt $\gamma$-ray phase) can account for the absence of the
IR/optical flashes in GRB afterglows. Again, before accepting such
an interpretation, we need independent probe of the physical
composition of the GRB outflow. Anyway, we do find that in some
bursts, for example, GRB 050319 \cite{Mason06}, GRB 050401
\cite{Rykoff05}, and GRB 061007 \cite{Mundell07,Schady07}, the
optical afterglow flux drops with time as a single power for $t>
{\rm several}\times 100$ s and strongly implies a very small
$\nu_m(t_\times)$. It is likely that the non-detection of the
IR/optical flash in some bursts are consistent with the standard
reverse shock model and thus not to our surprise.

So far the physical composition of the GRB outflow is not clear,
yet. If the GRB outflow is just mildly magnetized, there are two
important signatures. One is that the prompt emission as well as
the reverse shock emission should be highly polarized (Lyutikov,
Pariev  \& Blandford 2003; Granot 2003; Fan et al. 2004). The
other is that the reverse shock emission should be much brighter
than the non-magnetization case \cite{Fan04,ZK05}. It is
interesting to note that both signatures may have been detected in
GRB 041219a.  By modelling the reverse/forward shock emission of
GRB 041219a, Fan et al. (2005b) showed that the reverse shock
region was weakly magnetized. Very recently, McGlynn et al. (2007)
found possible evidence for the high linear polarization of the
prompt $\gamma-$rays. These findings are consistent with the
mildly magnetized outflow model of GRBs. Thanks to the successful
performance of the ring polarimeter on the robotic Liverpool
Telescope, a reliable UV/optical polarimetry of the prompt or
reverse shock emission is possible \cite{Mundell07a}. The nature
of the outflow could thus be better constrained in the near
future.


\section*{Acknowledgments}
We thank Bing Zhang and Daming Wei for discussion, and Daniele
Malesani, Dong Xu and Ting Yan for kind help. We also appreciate
the referee for his/her constructive suggestions. This work is
supported by the National Natural Science Foundation (grant
10673034) of China and by a special grant of Chinese Academy of
Sciences.


\begin{thebibliography}{99}
\bibitem[]{} Akerlof C., et al. 1999, Nature, 398, 400
\bibitem[Blake et al. 2005]{Blake05} Blake C. H., et al., 2005,
Nature, 435, 181
\bibitem[Blandford \& McKee 1976]{BM76} Blandford R. D., McKee C. F., 1976, Phys. Fluids., 19, 1130
\bibitem[Bo\"{e}r \etal 2006]{Boer06} Bo\"er, M., Atteia, J. L., Damerdji, Y., Gendre, B., Klotz, A., \&
Stratta, G. 2006, ApJ, 638, L71
\bibitem[Chevalier \& Li 2000]{CL00} Chevalier R. A., Li Z. Y., 2000, ApJ, 536, 195
\bibitem[Dai \& Lu 1998]{DL98} Dai Z. G., Lu T., 1998, MNRAS, 298, 87
\bibitem[Fan et al. 2002]{Fan02} Fan Y. Z., Dai Z. G., Huang Y. F., Lu T., 2002,
Chin. J. Astron. Astrophys., 2, 449
\bibitem[Fan \& Piran 2006a]{FP06a} Fan Y. Z., Piran T., 2006a,
MNRAS, 369, 197
\bibitem[Fan \& Piran 2006b]{FP06} Fan Y. Z., Piran T., 2006b,
MNRAS, 370, L24
\bibitem[Fan \etal 2006]{FPX06} Fan Y. Z., Piran T., Xu D., 2006, JCAP,
0609, 013
\bibitem[Fan \& Wei 2005]{FW05} Fan Y. Z., Wei D. M., 2005, MNRAS, 364, L42
\bibitem[Fan \etal 2004]{Fan04} Fan Y. Z., Wei D. M., Wang C. F., 2004, A\&A, 424,
477
\bibitem[Fan \etal 2007]{FWX07} Fan Y. Z., Wei D. M., Xu D., 2007, MNRAS in press
(astro-ph/0702006)
\bibitem[Fan, Zhang \& Proga 2005a]{FZP05} Fan Y. Z., Zhang B., Proga D., 2005a, ApJ, 635, L129
\bibitem[Fan, Zhang \& Wei 2005b]{FZW05} Fan Y. Z., Zhang B., Wei D. M., 2005b, ApJ, 628,
L25
\bibitem[Gao \& Fan 2006]{GF06} Gao W. H., Fan Y. Z., 2006, Chin. J. Astron. Astrophys,
              6, 513
\bibitem[Gorosabel et al. 2006]{Goro06} Gorosabel J., et al.,
2006, A\&A, 459, L33
\bibitem[Granot 2003]{G03} Granot, J. 2003, ApJ, 596, L17
\bibitem[Katz \etal 1998]{KPS98} {Katz J. I., Piran T., Sari R.}, 1998, {Phys. Rev.
Lett.}, {80}, {1580}
\bibitem[Kennel \& Coronitti 1984]{KC84} Kennel C. F., \& Coronitti E. V., 1984, ApJ, 283, 694
\bibitem[Kobayashi 2000]{Kobayashi00} Kobayashi S., 2000, ApJ, 545, 807
\bibitem[Lyutikov \etal 2003]{LPB03} Lyutikov, M., Pariev, V. I., \& Blandford, R. D. 2003,
ApJ, 597, 998
\bibitem[Mason \etal 2006]{Mason06} Mason K. O., et al., 2006, ApJ,
639, 361
\bibitem[McGlynn et al. 2007]{McGl07}
McGlynn S \etal, 2007, A\&A, in press (astro-ph/0702738)
\bibitem[McMahon \etal 2006]{MKP06} McMahon E., Kumar P., Piran,
T., 2006, MNRAS, 366, 575
\bibitem[M\'esz\'aros \& Rees 1999]{MR99} M\'esz\'aros, P., \& Rees, M. J., 1999, MNRAS, 306, L39
\bibitem[Molinari \etal 2007]{Molinari07} Molinari E. \etal, 2007, A\&A, submitted (astro-ph/0612607)
\bibitem[Mundell \etal 2007a]{Mundell07a} Mundell C. G. et al.,
2007a, Science, in press (astro-ph/0703654)
\bibitem[Mundell \etal 2007b]{Mundell07} Mundell C. G. et al., 2007b, ApJ,
in press (astro-ph/0610660)
\bibitem[Nousek \etal 2006]{Nousek06} {Nousek, J. A. et al.,} {2006}, {\apj}, {642}, {389}
\bibitem[Nakar \& Piran 2004]{NP04} Nakar E., Piran T., 2004,
MNRAS, 353, 647
\bibitem[Panaitescu \& Kumar 2001]{PK01} Panaitescu A., Kumar P., 2001, \apj, 560, 49
\bibitem[Piran 1999]{P99} Piran T., 1999, Phys. Rep., 314, 575
\bibitem[Ramirez-Ruiz \etal 2001]{RR01}
 Ramirez-Ruiz E, Dray  L. M., Madau P., Tout C. A.,
 2001, MNRAS, 327, 829
\bibitem[Roming \etal 2006]{Roming06} Roming, P. W. A., \etal,
2006 , ApJ, 652, 1416
\bibitem[Rykoff \etal 2005]{Rykoff05} Rykoff E. S., et al. 2005, ApJ, 631, L121
\bibitem[Sari \& Piran 1999a]{SP99} Sari R.,  Piran T., 1999a, ApJ, 517, L109
\bibitem[Sari \& Piran 1999b]{SP99b} Sari R.,  Piran T., 1999b, ApJ, 520,
641
\bibitem[Sari \etal 1998]{SPN98} Sari R., Piran T., Narayan R., 1998, ApJ, 497, L17
\bibitem[Schady \etal 2007]{Schady07} Schady P., 2007, MNRAS,
submittd (astro-ph/0611081)
\bibitem[Sollerman \etal 2007]{Sollerman07} Sollerman J., 2007,
A\&A, in press (astro-ph/0701736)
\bibitem[Troja \etal 2007]{Troja07} Troja E. \etal, 2007,
ApJ, submitted (astro-ph/070220)
\bibitem[Vestrand \etal 2005]{Vest05} Vestrand W. T. et al., 2005,
Nature, 435, 178
\bibitem[Wu \etal 2007]{Wu07} Wu X. F., Dai Z. G., Wang X. Y., Huang Y. F., Feng L. L., Lu T.,
 2007, ApJ, submitted (astro-ph/0512555)
\bibitem[Yan \etal 2007]{Yan06} Yan T., Wei D. M., Fan Y. Z.,
2007, Chin. J. Astron. Astrophys., submitted (astro-ph/0512179)
\bibitem[Zhang 2007]{ZhangB07} Zhang B., 2007, Chin. J. Astron.
Astrophys., 7, 1
\bibitem[Zhang \etal 2006]{Z06} Zhang B., Fan Y. Z., Dyks J., Kobayashi S., M\'esz\'aros P.,
Burrows D. N., Nousek J. A., Gehrels N.  2006, ApJ, 642, 354
\bibitem[Zhang \& Kobayashi 2005]{ZK05} Zhang B., Kobayashi S., 2005, ApJ,
628, 315
\bibitem[Zhang \etal 2003]{Zhang03} Zhang B., Kobayashi S., M\'esz\'aros P., 2003, ApJ,
595, 950
\bibitem[Zhang et al. 2007]{Zhang07} Zhang B. B., Liang E. W.,
\& Zhang B., ApJ, submitted (astro-ph/0612246)
\end{thebibliography}
\end{document}